\documentclass[twocolumn, english]{revtex4}
\usepackage[T1]{fontenc}
\usepackage[latin9]{inputenc}
\usepackage{amsmath}
\usepackage{graphicx}
\usepackage{amssymb}

\makeatletter
\@ifundefined{textcolor}{}
{%
 \definecolor{BLACK}{gray}{0}
 \definecolor{WHITE}{gray}{1}
 \definecolor{RED}{rgb}{1,0,0}
 \definecolor{GREEN}{rgb}{0,1,0}
 \definecolor{BLUE}{rgb}{0,0,1}
 \definecolor{CYAN}{cmyk}{1,0,0,0}
 \definecolor{MAGENTA}{cmyk}{0,1,0,0}
 \definecolor{YELLOW}{cmyk}{0,0,1,0}
 }

\makeatother

\makeatother

\usepackage{babel}

\begin{document}

\title{Enhanced relativistic harmonics by electron nanobunching}

\author{D. an der Br\"ugge}

\email{dadb@tp1.uni-duesseldorf.de}

\author{A. Pukhov}

\affiliation{Institut für theoretische Physik I, Heinrich-Heine-Universit\"at D\"usseldorf }
\begin{abstract}
It is shown that when an few-cycle, relativistically intense, $p$-polarized
laser pulse is obliquely incident on overdense plasma, the surface
electrons may form ultra-thin, highly compressed layers, with a width
of a few nanometers. These electron {}``nanobunches'' emit synchrotron
radiation coherently. We calculate the one-dimensional synchrotron
spectrum analytically and obtain a slowly decaying power-law with
an exponent of $4/3$ or $6/5$. This is much flatter than the $8/3$
power of the BGP (Baeva-Gordienko-Pukhov) spectrum, produced by a
relativistically oscillating bulk skin layer. The synchrotron spectrum
cut-off frequency is defined either by the electron relativistic $\gamma$-factor,
or by the thickness of the emitting layer. In the numerically demonstrated,
locally optimal case, the radiation is emitted in the form of a single
attosecond pulse, which contains almost the entire energy of the full optical cycle.
\end{abstract}
\maketitle

\section{Introduction}

High harmonics generated by an intense laser pulse incident on an
overdense plasma surface are a promising bright source of short wavelength
radiation with a number of potential applications \cite{PukhovNatPhys2006}.
Because of the coherent character of the process, the generated harmonics
are locked in phase and emerge in the form of attosecond pulses. This
could recently also be shown experimentally \cite{NomuraNatPhys2009}.
Commonly, two mechanisms that lead to harmonics generation from solid
density plasma surfaces are distinguished: coherent wake emission
(CWE) \cite{quere:125004} and the {}``relativistically oscillating
mirror'' (ROM) \cite{lichters:3425}.

CWE is predominant at moderately relativistic intensities $a_{0}\sim1$
and short, but finite plasma gradient lengths $\nabla N/N\sim0.01\,\lambda$.
Here, $a_{0}=eA/mc^{2}$ is the relativistically normalized laser
potential. In this regime the Brunel electrons that re-enter the plasma
excite electrostatic oscillations in the overcritical plasma regions.
Because of the strong density inhomogeneity, the electrostatic oscillations
couple back to electromagnetic modes and thus generate harmonics.
Due to this mechanism, CWE spectra have a cut-off at the plasma frequency
corresponding to the maximum density. Further, the harmonics have
a comparatively wide emission cone \cite{Dromey2009}.

ROM harmonics dominate for $a_{0}\gg1$, although recent studies \cite{tarasevitch:103902}
have shown that they can be observed even at lower intensities when
the plasma gradient length is about $\nabla N/N\sim0.1\,\lambda$.
ROM spectra have no cut-off at the plasma frequency, but can extend
to much higher frequencies. Their origin is basically the relativistic
non-linearity in the electron motion. Baeva, Gordienko and Pukhov
(BGP) \cite{bgp-theory2006} have developed a theory to analytically
describe the spectrum in this highly relativistic regime. The BGP
harmonic spectrum is a power law with the exponent $-8/3$, smoothly
rolling off into an exponential decay at some frequency scaling as
$I^{3/2}$. This type of spectrum could be observed experimentally
\cite{Dromey2006}.

So far these two mechanisms have been the widely accepted explanations
for harmonics generation at overdense plasma surfaces. However, there
has also been some evidence that the two models do not tell the whole
story. Especially under $p$-polarized oblique incidence, with $a_{0}\gg1$
some spectra could be observed in numerical simulations \cite{boyd:125004}
that do not fit to the predictions of neither of the models. 

In this paper, we take the most general approach to describe the harmonics
radiation of the relativistic electrons: It is described as coherent
synchrotron emission of the one dimensional slab geometry electron
distribution. For synchrotron radiation, the optimal efficiency is
reached, when all the radiating electrons gather in one extremely
dense and narrow bunch. We demonstrate via PIC simulation, that this
case can actually closely be achieved in relativistic laser-plasma
interaction.

\section{Assumptions and Predictions of BGP Theory}

We start with a brief review of the BGP theory. The theory is one
dimensional, oblique incidence can be treated in a Lorentz transformed
frame \cite{Bourdier1983}. This is acceptable as long as the focal
spot size is big compared to the laser wavelength \cite{adBPuk2007}.
The reflection of the light is done by the electrons close to the
surface. In general, the surface electrons perform highly complicated
motions and it is a hopeless endeavour to try to describe it analytically
even in 1D. Therefore we have to resort to some approximation. The starting
point of the BGP model is the following, simplified boundary condition:

\begin{equation}
E_{i}\left(t-x_{\textrm{ARP}}(t)/c\right)+E_{r}\left(t+x_{\textrm{ARP}}(t)/c\right)=0\label{eq:ARP_boundary}\end{equation}

\noindent where $x_{\textrm{ARP}}$ is the so called {}``Apparent
Reflection Point'' (ARP) position (see Eq.~(18) in \cite{bgp-theory2006}).
The whole information about the plasma motion has been condensed to
only one real function $x_{\textrm{ARP}}(t)$. However, we have to
emphasize here, that Eq.~\eqref{eq:ARP_boundary} is an approximation.
Although it is not yet possible to give a fully general criterion
for the validity of Eq.~\eqref{eq:ARP_boundary} in terms of the initial
laser and plasma parameters, it has been demonstrated 
in \cite{PhysRevLett.93.115002} that the approximation
is valid if the plasma skin layer evolution time $\tau$ is long compared
to the skin length $\delta$ in the sense $c\tau\gg\delta$ . This
is the case if $S\equiv N_{e}/a_{0}N_{c}\gg1$, or generally if the
electron density profile is approximately step-like during the main
interaction phase.

\noindent Further note that Eq.~\eqref{eq:ARP_boundary} makes sense
only for an ARP moving at a physical velocity of $|\dot{x}_{\textrm{ARP}}|<c$.
In this case it is clear from Eq.~\eqref{eq:ARP_boundary}, that
the reflected field is nothing but a phase modulation of the negative
of the incident field. This can easily be checked inside a PIC simulation.
Fig.~\ref{fig:time_domain+spectra}a shows an exemplary result. For
this simulation, the laser was normally incident on a sharp edged
plasma density profile. Here, Eq.~\eqref{eq:ARP_boundary} is fulfilled
to a good approximation as can be verified from Fig.~\ref{fig:time_domain+spectra}a:
Apart from minor deviations, the extremal values and the sequence
of the monotonic intervals agree, the reflected field is approximately
a phase modulation of the incident one.

\noindent Starting from Eq.~\eqref{eq:ARP_boundary}, it is now possible
to calculate the spectrum of the reflected radiation with only little
knowledge of the function $x_{\textrm{ARP}}(t)$. Therefore we perform
the Fourier transformation of $E_{r}(t)$ employing the method of
stationary phase. At the stationary phase point, the relativistic
$\gamma$-factor corresponding to the ARP $\gamma_{\textrm{ARP}}=(1-(\min(\dot{x}_{\textrm{ARP}}/c))^{2})^{-1/2}$
possesses an extremely sharp maximum, why we also refer to this point
as the $\gamma$-spike. The high frequency spectrum does only depend
on the surface behavior near this $\gamma$-spike. Therefore, we can
also relax the condition for the validity of the theory: Actually,
we require Eq.~\eqref{eq:ARP_boundary} only to be fulfilled in the
neighbourhood of the $\gamma$-spike. 

\noindent The exact result of the calculation of the spectrum is given
by Eqs.~(34)-(36) in \cite{bgp-theory2006}, in a very good approximation
it can be simplified to:

\begin{equation}
I(\omega)\propto\frac{1}{\omega^{8/3}}\,\left[\textrm{Ai}\left(\left(\frac{\omega}{\omega_{r}}\right)^{2/3}\right)\right]^{2}\label{eq:ARP_spec}\end{equation}

\noindent wherein $\textrm{Ai}$ refers to the well known Airy-function.
The roll-off frequency $\omega_{r}$ marks the point, where the initial
power law decay of the spectral envelope merges into a more rapid
exponential decay. In the region $\omega\ll\omega_{r}$, the Airy
function is $O(1)$ and one obtains the $-8/3$-power law spectrum.
For $\omega\gg\omega_{r}$, the decay is an exponential one. From
the integration, it also follows that $\omega_{r}\propto\gamma_{\textrm{ARP}}^{3}$,
where $\gamma_{\textrm{ARP}}$ is taken at its maximum. The scaling
with $\gamma_{\textrm{ARP}}^{3}$ is in contrast to the long known
case of reflection at a mirror moving with constant velocity generates
a Doppler upshift of only $(1+\beta)/(1-\beta)\approx4\gamma^{2}$.

Fig.~\ref{fig:time_domain+spectra}b shows the spectrum corresponding
to the field from Fig.~\ref{fig:time_domain+spectra}a. As expected,
it agrees well with the $-8/3$-power law.

\section{\label{sec:New-Regime}A New Regime of Relativistic Harmonics Generation}

\begin{figure}
\includegraphics[width=1\columnwidth]{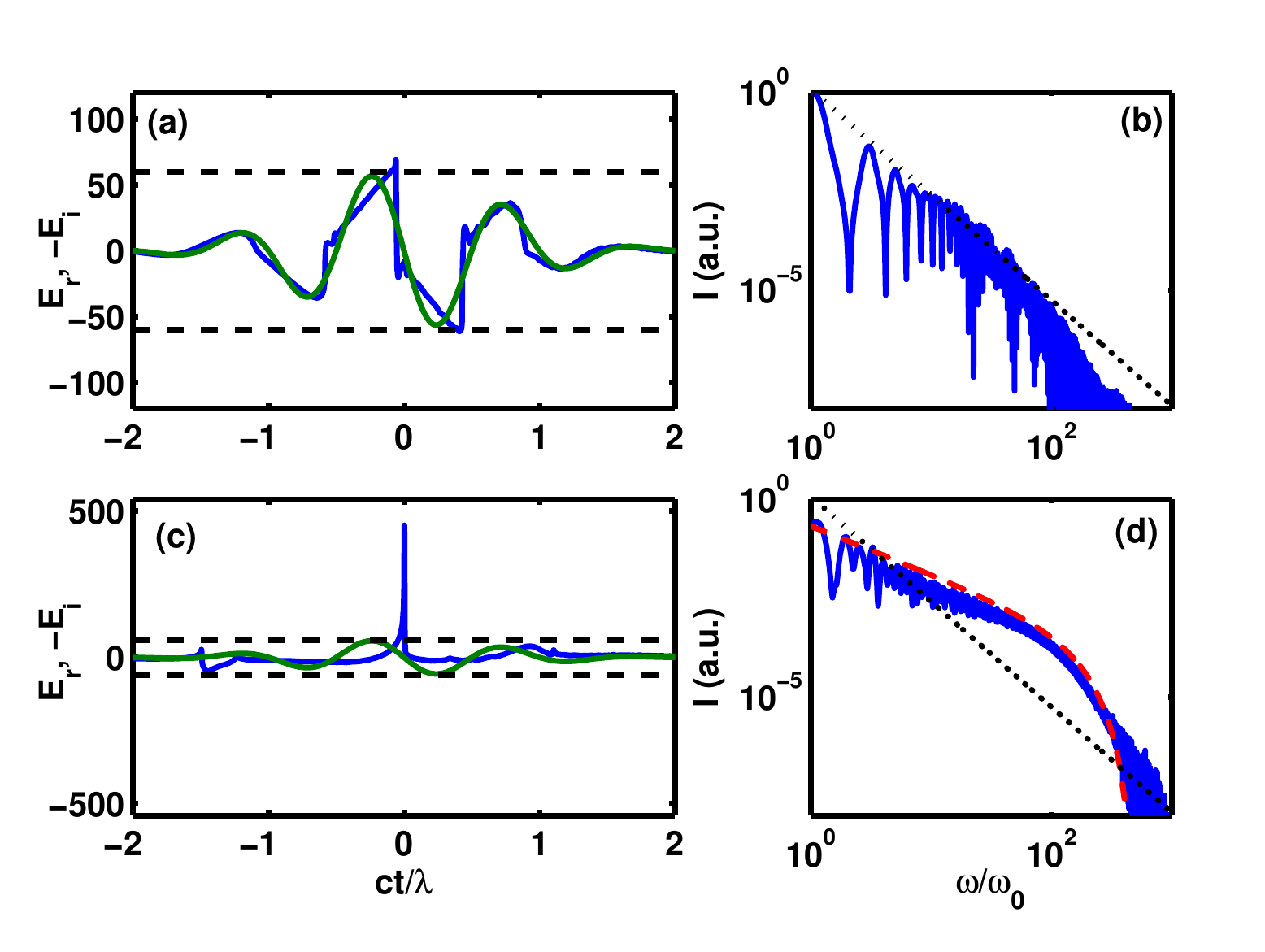}

\caption{\label{fig:time_domain+spectra}Radiation in time ((a) and (c)) and
spectral ((b) and (d)) domain for two simulations. (a) and (b) correspond
to the BGP case: normal incidence, plasma density $N_{e}=250\, N_{c}$,
sharp edged profile; (c) and (d) correspond to the nanobunching case:
plasma density ramp $\propto\exp(x/(0.33\,\lambda))$ up to a maximum
density of $N_{e}=95\, N_{c}$ (lab frame), oblique incidence at $63^{\circ}$
angle ($p$-polarized). Laser field amplitude is $a_{0}=60$ in both
cases. In all frames, the reflected field is represented by a blue
line. In (a) and (c), the green line represents the field of the incident
laser and the black dashed lines mark the maximum field of it. In
(b) and (d), the dotted black line represents an $8/3$ power law,
the red dashed line corresponds to the 1D synchrotron spectrum Eqs.~\eqref{eq:synch_spec_2nd}
and \eqref{eq:gaussian_shape}, with $\omega_{rs}=800\,\omega_{0}$
and $\omega_{rf}=225\,\omega_{0}$.}

\end{figure}

Let us now have a look at Fig.~\ref{fig:time_domain+spectra}c, which
shows the result of another PIC simulation. In this simulation the
plasma density profile was extended over a few fractions of a laser
wavelength, and the $p$-polarized laser was incident at an angle
of $\alpha=63^{\circ}$. The laser pulse used was the same as in the
first simulation.

It is evident, that the maximum of the reflected field reaches out
about an order of magnitude above the amplitude of the incident laser.
The reflected radiation can clearly \textit{not} be obtained from
the incident one just by phase modulation. By this we conclude, that
the boundary condition Eq.~\eqref{eq:ARP_boundary} fails here. Consequently,
the spectrum deviates from the $-8/3$-power law, compare Fig.~\ref{fig:time_domain+spectra}d.
Indeed we see that the efficiency of harmonic generation is much higher
than estimated by the BGP calculations: about two orders of magnitude
at the hundredth harmonic. Also, we can securely exclude CWE as the
responsible mechanism, since this would request a cut-off around $\omega=10\omega_{0}$.
From this we conclude that the radiation cannot be attributed to any
of the known mechanisms.

\begin{figure}
\includegraphics[width=1\columnwidth]{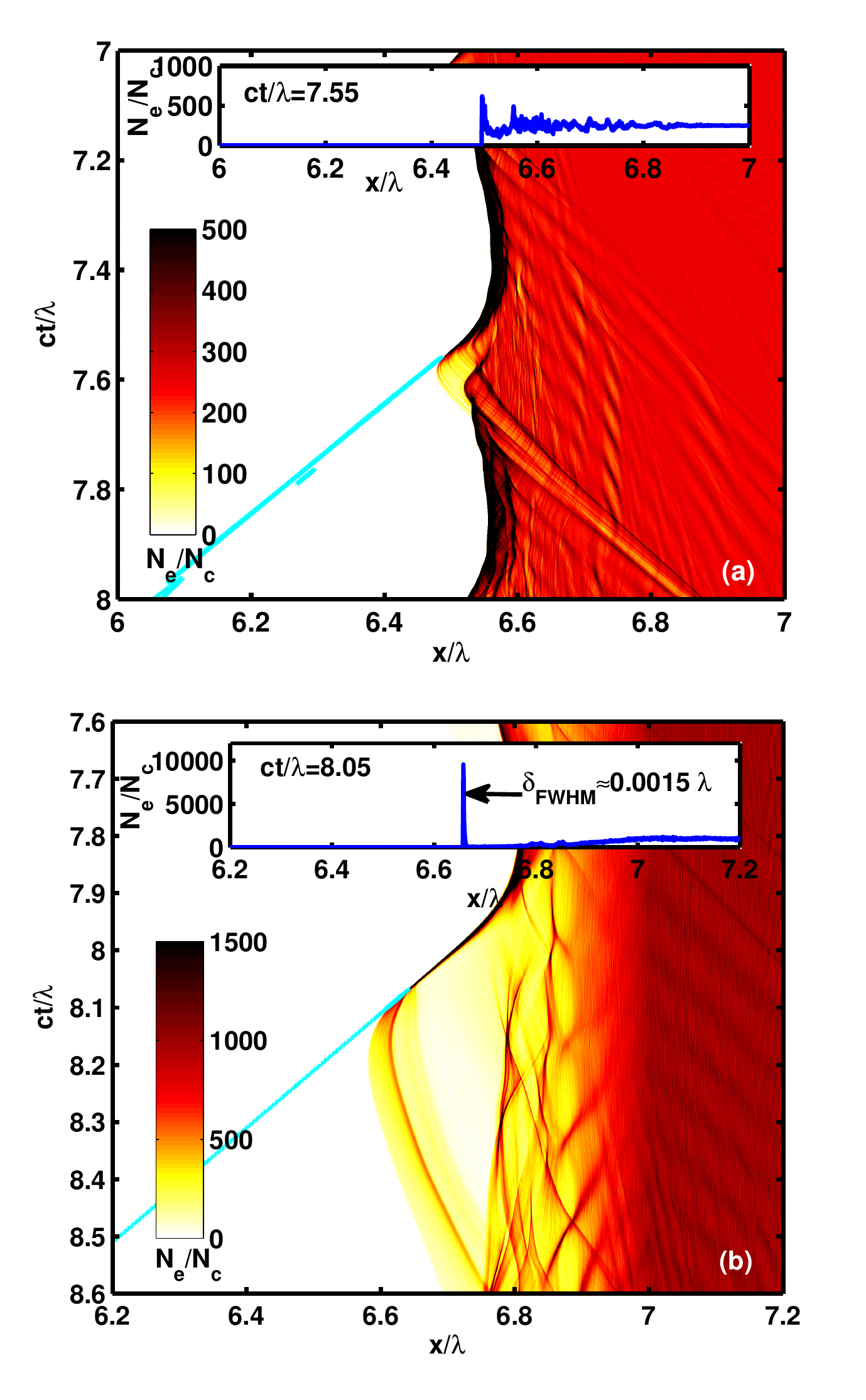} 

\caption{\label{fig:generation}(Color online) The electron density and contour
lines (cyan, or light grey) of the emitted harmonics radiation for
$\omega/\omega_{0}>4.5$, in (a) the BGP and (b) the synchrotron (or
nanobunching) regime. The small windows inside the main figures show
the detailed density profile at the instant of harmonic generation.
All magnitudes are taken in the simulation frame. The parameters used
are the same as in the other figures.}

\end{figure}

To get a picture of the physics behind, let us have a look at the
motion of the plasma electrons that generate the radiation. Figure~\ref{fig:generation}
shows the evolution of the electron density in both our sample cases.
In addition to the density, contour lines of the spectrally filtered
reflected radiation are plotted. These lines illustrate where the
main part of the high frequency radiation emerges. We observe that
in both cases the main part of the harmonics is generated at the point,
when the electrons move towards the observer. This shows again that
in both cases the radiation does not stem from CWE. For CWE harmonics,
the radiation is generated inside the plasma, at the instant when
the Brunel electrons re-enter the plasma \cite{quere:125004}. Apart
from that mutuality, the two presented cases appear to be very different.

Figure~\ref{fig:generation}a corresponds to the BGP case. It can
be seen that the density profile remains roughly step-like during
the whole interaction process and the plasma skin layer radiates as
a whole. This explains why the BGP theory works well here, as we have
seen before in Fig.~\ref{fig:time_domain+spectra}a and b.

However, figure~\ref{fig:generation}b looks clearly different. The
density distribution at the moment of harmonics generation is far
from being step-like, but possesses a highly dense (up to $\sim10000\, N_{c}$
density) and very narrow $\delta$-like peak, with a width of only
a few nanometers. This electron {}``nanobunch'' emits synchrotron
radiation coherently.

\begin{figure}

\includegraphics[width=1\columnwidth]{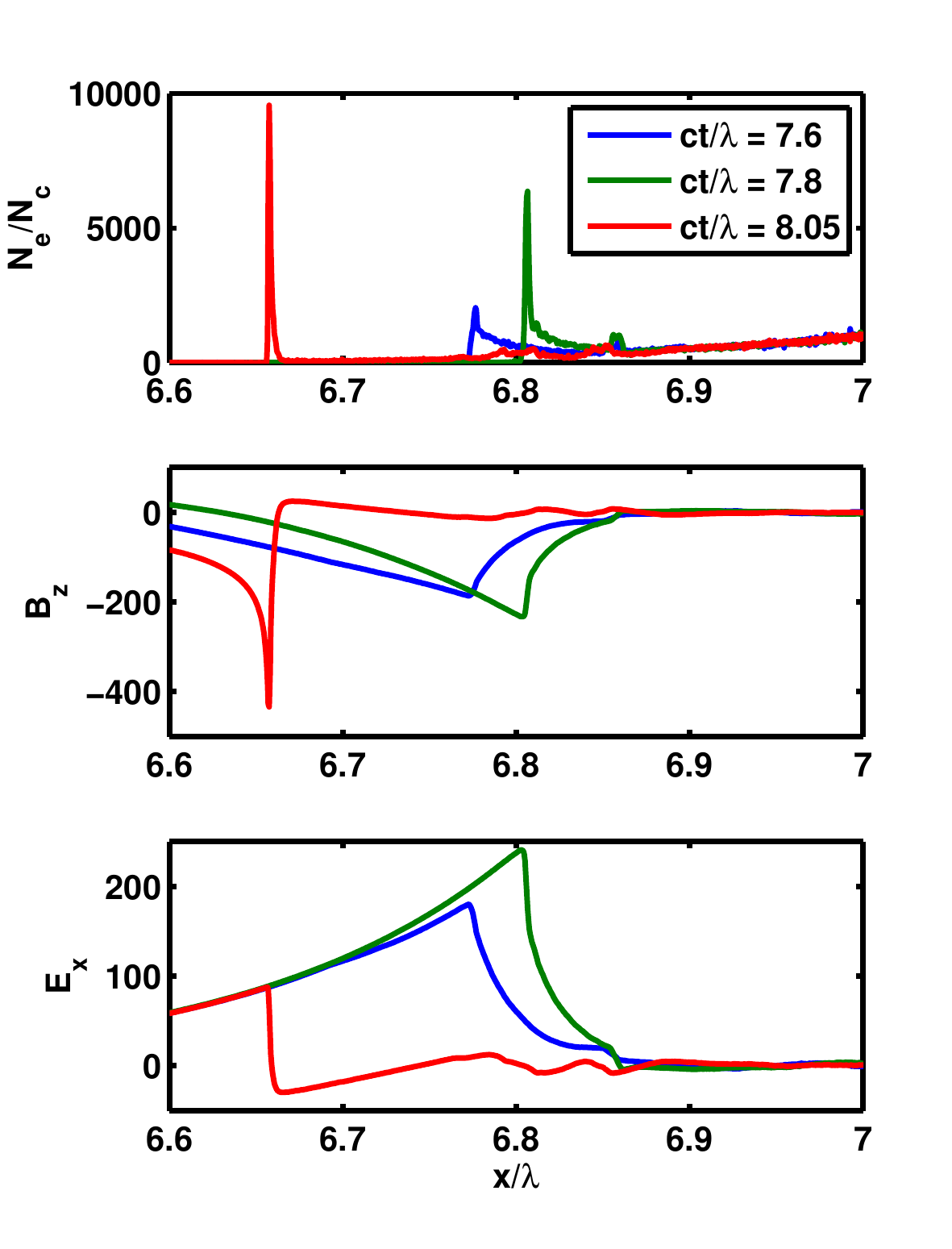}\caption{\label{fig:formation}(Color online) Formation of the nanobunch in
the simulation corresponding to Figs.~\ref{fig:time_domain+spectra}c-d
and \ref{fig:generation}b. We depict the electron density $N_{e}$
in units of the critical density $N_{c}$, the transverse magnetic
field component $B_{z}$ and the longitudinal electric field component
$E_{x}$ in relativistically normalized units.}

\end{figure}

The radiation is emitted by a highly compressed electron bunch moving
\emph{away} from the plasma. However, the electrons first become compressed
by the relativistic ponderomotive force of the laser that is directed
into the plasma, compare the blue lines in Fig.~\ref{fig:formation}.
During that phase, the longitudinal electric field component grows
until the electrostatic force turns around the bunch, compare the
green lines in Fig.~\ref{fig:formation}. Normally, the bunch will
loose its compression in that instant, but in some cases, as in the
one considered here, the fields and the bunch current match in a way
that the bunch maintains or even increases its compression. The final
stage is depicted by the red lines in Fig.~\ref{fig:formation}.

We emphasize, that such extreme nanobunching does not occur in every
case of $p$-polarized oblique incidence of a highly relativistic
laser on an overdense plasma surface. On the contrary, it turns out
that the process is highly sensitive to changes in the plasma density
profile, laser pulse amplitude, pulse duration, angle of incidence
and even the carrier envelope phase of the laser. For a longer pulse, we
may even observe the case, that nanobunching is present in some optical cycles but not in others. The parameters in
the example were selected in a way to demonstrate the new effect unambiguously,
i.e. the nanobunch is well formed and emits a spectrum that clearly
differs from the BGP one. The dependence of the effect on some parameters
is discussed in section~\ref{sec:Parametrical-Dependence}.

Because of the one dimensional slab geometry, the spectrum is not
the same as the well known synchrotron spectra \cite{Jackson} of
a point particle. We now calculate the spectrum analytically.

\section{\label{sec:CSE}Spectrum of 1D Coherent Synchrotron Emission (CSE)}

The radiation field generated to the left of a one-dimensional current
distribution can in a completely general way be expressed as:

\begin{equation}
E_{sy}\left(t,x\right)=\frac{2\pi}{c}\int_{-\infty}^{+\infty}j\left(t+\frac{x-x'}{c},x'\right)\, dx'\label{eq:synch_field}\end{equation}

\noindent Optimal coherency for high frequencies will certainly be
achieved, if the current layer is infinitely narrow: $j(t,x)=j(t)\delta(x-x_{el}(t))$.
To include more realistic cases, we allow in our calculations for
a narrow, but finite electron distribution: \begin{equation}
j(t,x)=j(t)f(x-x_{el}(t))\label{eq:current}\end{equation}
with variable current $j(t)$ and position $x_{el}(t)$, but fixed
shape $f(x)$. We take the Fourier transform of Eq.~\eqref{eq:synch_field},
thereby considering the retarded time, and arrive at the integral
$\tilde{E}_{sy}(\omega)=2\pi c^{-1}\tilde{f}(\omega)\int j(t)\,\exp\left[-i\omega\left(t+x_{el}(t)/c\right)\right]\, dt$.
Here, $\tilde{f}(\omega)$ denotes the Fourier transform of the shape
function. In analogy to the standard synchrotron radiation by a point
particle, the integral can be solved with the method of stationary
phase. Therefore we note, that for high $\omega$ the main contributions
to the integral come from the regions, where the phase $\Phi=\omega\left(t+x_{el}(t)/c\right)$
is approximately stationary, i.e. $d\Phi/dt\approx0$. However, to
get some kind of result we need some assumption about the relation
between the functions $j(t)$ and $x_{el}(t)$. Since we are dealing
with the ultrarelativistic regime $a_{0}\gg1$, it is reasonable to
assume that changes in the velocity components are governed by changes
in the direction of movement rather than by changes in the absolute
velocity, which is constantly very close to the speed of light $c$.

\noindent Let the electron motion in momentum space be given by $\left(p_{x},p_{y}\right)=a_{0}m_{e}c\left(\hat{p}_{x}\,\hat{p}_{y}\right)$,
so that $j(t)=ecn_{e}\hat{p}_{y}/(a_{0}^{-2}+\hat{p}_{x}^{2}+\hat{p}_{y}^{2})$
and $\dot{x}_{el}=c\hat{p}_{x}/(a_{0}^{-2}+\hat{p}_{x}^{2}+\hat{p}_{y}^{2})$.
The derivative of the phase approaches zero at points where $\dot{x}_{el}\approx-c$,
so from our assumption of ultrarelativistic motion we conclude $\hat{p}_{y}=0$
at these instants. Now, two cases have to be distinguished:
\begin{enumerate}
\item \noindent The current changes sign at the stationary phase point.
Then we can Taylor expand $j(t)=\alpha_{0}\, t$ and $x_{el}(t)=-v_{0}t+\alpha_{1}t^{3}/3$.
The integral can now be expressed in terms of the well-known Airy-function,
yielding $\tilde{E}_{sy}(\omega)=C\,\tilde{f}(\omega)\,\omega^{-2/3}\textrm{Ai}'\left(\frac{1-v_{0}}{\sqrt[3]{\alpha_{1}}}\omega^{2/3}\right),$
where $\textrm{Ai}'$ is the Airy function derivative and $C=i(2\pi)^{2}en_{e}c^{-1}\alpha_{0}\alpha_{1}^{-2/3}$
is a complex prefactor. We now find the spectral envelope\begin{equation}
I(\omega)\propto|\tilde{f}(\omega)|^{2}\,\omega^{-4/3}\,\left[\textrm{Ai}'\left(\left(\frac{\omega}{\omega_{rs}}\right)^{2/3}\right)\right]^{2}\label{eq:synch_spec_1st}\end{equation}
with $\omega_{rs}\approx2^{3/2}\sqrt{\alpha_{1}}\gamma_{0}^{3}$,
where $\gamma_{0}=(1-v_{0}^{2})^{-1/2}$ is the relativistic $\gamma$-factor
of the electron bunch at the instant when the bunch moves towards
the observer. As in the BGP case, the spectral envelope \eqref{eq:synch_spec_1st}
does not depend on all details of the electron bunch motion $x_{el}$,
but only on its behavior close to the stationary points, i.e. the
$\gamma$-spikes.
\item The current does not change sign at the stationary phase point. Because
of the assumption of highly relativistic motion the changes in absolute
velocity can again be neglected compared to the changes in direction,
and it follows that in this case the third derivative of $x_{el}$
is zero at the stationary phase point. Therefore, our Taylor expansions
look like $j(t)=\alpha_{0}t^{2}$ and $x_{el}(t)=-v_{0}t+\alpha_{1}t^{5}/5$.
This leads us to the spectral envelope\foreignlanguage{british}{ }\begin{equation}
I(\omega)\propto|\tilde{f}(\omega)|^{2}\,\omega^{-6/5}\,\left[\textrm{S}''\left(\left(\frac{\omega}{\omega_{rs}}\right)^{4/5}\right)\right]^{2}\label{eq:synch_spec_2nd}\end{equation}
with $\textrm{S}''$ being the second derivative of $\textrm{S}(x)\equiv(2\pi)^{-1}\int\exp\left[i\left(xt+t^{5}/5\right)\right]\, dt$,
a special case of the canonical swallowtail integral \cite{1984Swallowtail}.
For the characteristic frequency $\omega_{rs}$ we now obtain $\omega_{rs}\approx2^{5/4}\sqrt[4]{\alpha_{1}}\gamma_{0}^{2.5}$.
Because now even the derivative of $\ddot{x}_{el}$ is zero at the
stationary phase point, the influence of acceleration on the spectrum
decreases and the characteristic frequency scaling is closer to the
$\gamma^{2}$-scaling for a mirror moving with constant velocity.
\end{enumerate}
\begin{figure}
\includegraphics[width=0.8\columnwidth]{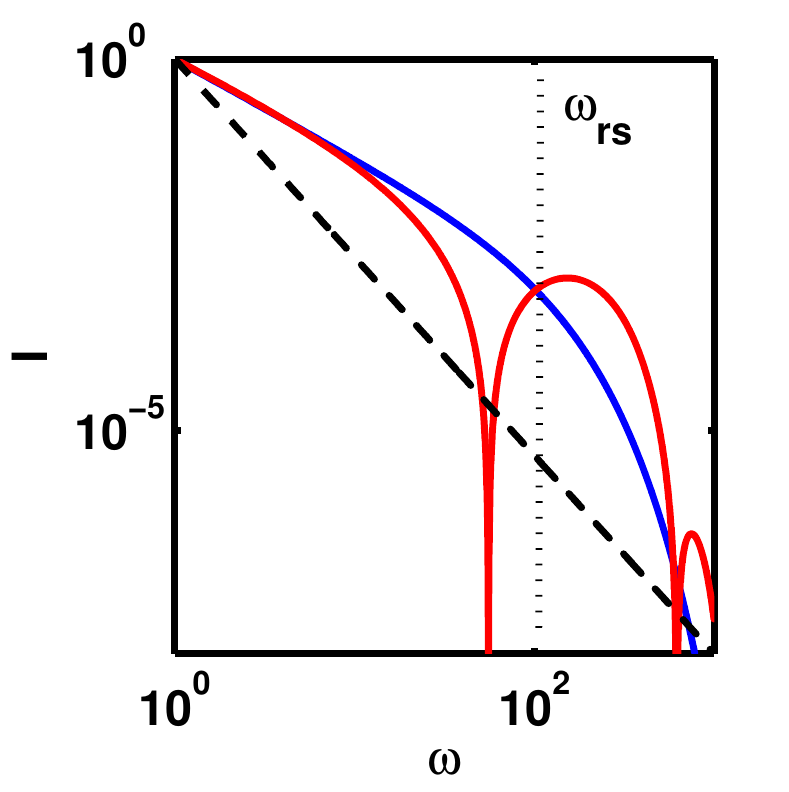}

\caption{\label{fig:CSE_spectra}(Color online) Coherent 1D synchrotron spectra
for an infinitely thin electron layer $\tilde{f}(\omega)\equiv1$
and $\omega_{rs}=100$. The blue line corresponds to Eq.~\eqref{eq:synch_spec_1st}
and the red line to Eq.~\eqref{eq:synch_spec_2nd}. For comparison,
the dashed black line denotes the BGP $8/3$-power law.}

\end{figure}

In Fig.~\ref{fig:CSE_spectra} the CSE spectra of the synchrotron
radiation from the electron sheets are depicted. Comparing them to
the $8/3$-power law from the BGP-case, we notice that, because of
the smaller exponents of their power law part, the CSE spectra are
much flatter, around the 100th harmonic we win more than two orders
of magnitude. Another intriguing property are the side maxima found
in the spectrum \eqref{eq:synch_spec_2nd}. This might provide an
explanation for modulations that are occasionally observed in harmonics
spectra, compare e.g. Ref.~\cite{boyd:125004}.

\begin{figure}
\includegraphics[width=0.8\columnwidth]{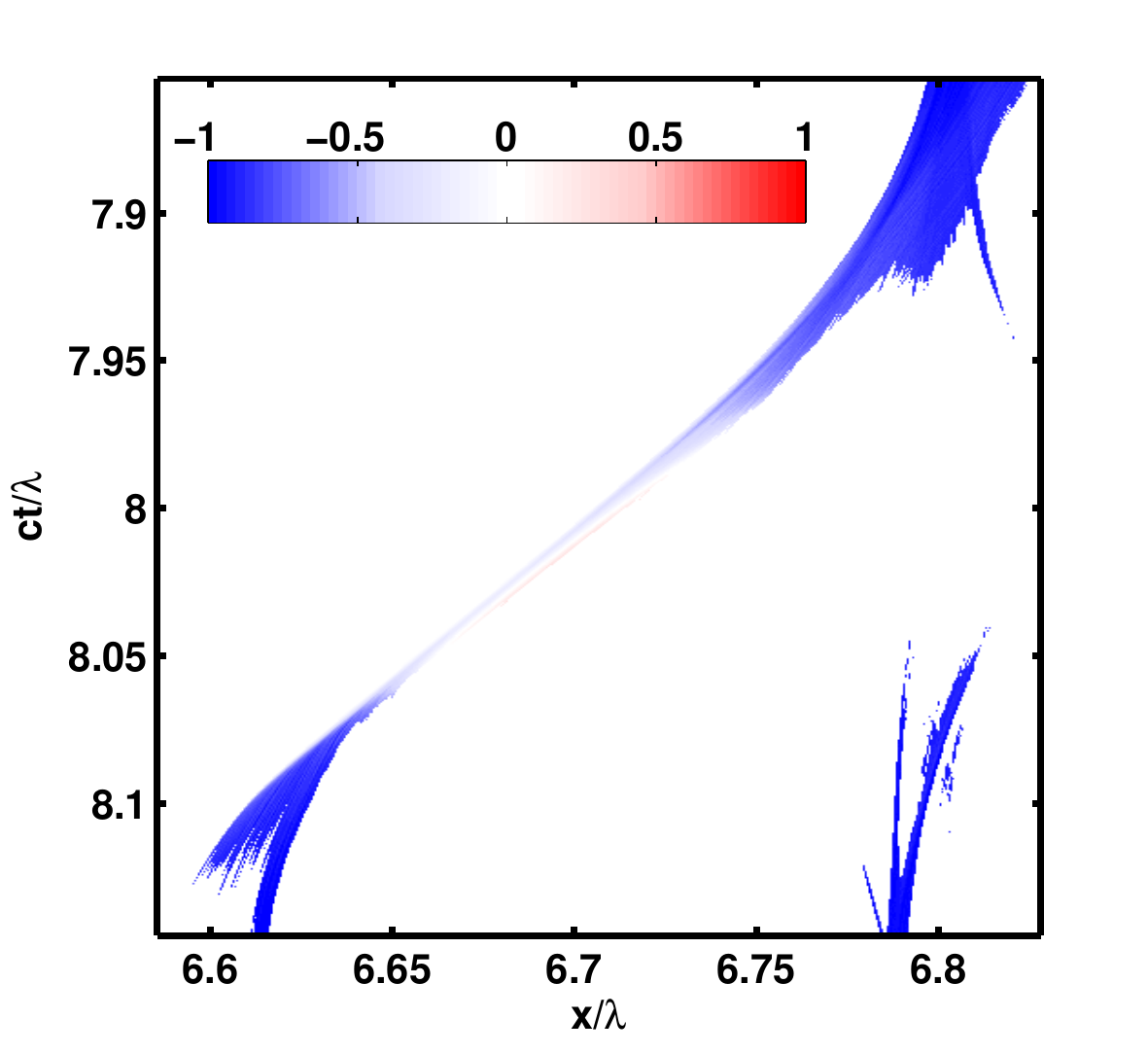}

\caption{\label{fig:vy}(Color online) Normalized transverse fluid velocity
$v_{y}/c$ of the electron nanobunch. Parts of the plasma with a density
below $500\, N_{c}$ are filtered out.}

\end{figure}

To compare with the PIC results, the finite size of the electron bunch
must be taken into account. Therefore we assume a Gaussian density
profile which leads us to 

\begin{equation}
|f(\omega)|^{2}=\exp\left[-\left(\frac{\omega}{\omega_{rf}}\right)^{2}\right]\label{eq:gaussian_shape}\end{equation}
Thus the spectral cut-off is determined either by $\omega_{rs}$,
corresponding to the relativistic $\gamma$-factor of the electrons,
or by $\omega_{rf}$ corresponding to the bunch width. A look at the
motion of the electron nanobunch in the PIC simulation (Fig.~\ref{fig:vy})
tells us that there is no change in sign of the transverse velocity
at the stationary phase point, consequently we use Eq.~\eqref{eq:synch_spec_2nd}.
We choose $\omega_{rf}=225\,\omega_{0}$ and $\omega_{rs}=800\,\omega_{0}$
to fit the PIC spectrum. $\omega_{rf}$ corresponds to a Gaussian electron 
bunch with a width of $\delta=10^{-3}\lambda$, which matches reasonably well to
the $\delta_{FWHM}=0.0015\,\lambda$ (see Fig.~\ref{fig:generation}b) measured
in the simulation, corresponding to a Gaussian electron bunch
$f(x)=\exp\left[-(x/\delta)^{2}\right]$ with a width of $\delta=10^{-3}\lambda$
and an energy of $\gamma\sim10$. This matches well with the measured
electron bunch width $\delta_{\text{FWHM}}=0.0015\,\lambda$ (see
figure~\ref{fig:generation}b) and the laser amplitude $a_{0}=60$, since we
expect $\gamma$ to be smaller but in the same order of magnitude as $a_0$.
In this case $\omega_{rf}<\omega_{rs}$, so the cut-off is dominated
by the finite bunch width. Still, both values are in the same order
of magnitude, so that the factor coming from the Swallowtail-function
cannot be neglected and actually contributes to the shape of the cut-off.
The modulations that appear in Fig.~\ref{fig:CSE_spectra} for frequencies
around $\omega_{rs}$ and above cannot be seen in the spectra, because
it is suppressed by the Gauss-function Eq.~\eqref{eq:gaussian_shape}.
The analytical synchrotron spectrum agrees excellently with the PIC
result, as the reader may verify in figure~\ref{fig:time_domain+spectra}d.

\section{Properties of the CSE Radiation}

As we see in Fig.~\ref{fig:time_domain+spectra}c, the CSE radiation
is emitted in the form of a single attosecond pulse whose amplitude
is significantly higher than that of the incident pulse. This pulse
has a FWHM duration of $0.003$ laser periods, i.e. $9\,\text{as}$
for a laser wavelength of $800\,\text{nm}$. This is very different
from emission of the ROM harmonics, which need to undergo diffraction
\cite{adBPuk2007} or spectral filtering \cite{bgp-theory2006} before
they take on the shape of attosecond pulses. 

When we apply spectral a spectral filter in a frequency range $\left(\omega_{\mathrm{low}},\omega_{{\scriptstyle \mathrm{high}}}\right)$
to a power-law harmonic spectrum with an exponent $q$, so that $I(\omega)=I_{0}(\omega_{0}/\omega)^{q}$
, the energy efficiency of the resulting attosecond pulse generation
process is 

\begin{eqnarray}
\eta_{\text{atto}} & = & \int_{\omega_{\mathrm{low}}}^{\omega_{\mathrm{high}}}I(\omega)\, d\omega\label{eq:cse_eff}\\
 & = & \frac{I_{0}\omega_{0}}{q-1}\left[\left(\frac{\omega_{0}}{\omega_{\text{low}}}\right)^{q-1}-\left(\frac{\omega_{0}}{\omega_{\text{high}}}\right)^{q-1}\right]\nonumber \end{eqnarray}
The scaling \eqref{eq:cse_eff} gives $\eta_{\text{atto}}^{\text{ROM}}\sim(\omega_{0}/\omega_{\text{low}})^{5/3}$
for the BGP spectrum with $q=8/3$. For unfiltered CSE harmonics with
the spectrum\textit{ $q=4/3$} the efficiency is close to $\eta_{\text{atto}}^{\text{CSE}}=1$.
This means that almost the whole energy of the original optical cycle
is concentrated in the attosecond pulse. Note that absorption is very
small in the PIC simulations shown; it amounts to 5\% in the run corresponding
to Fig.~\ref{fig:time_domain+spectra}c-d and is even less in the
run corresponding to Fig.~\ref{fig:time_domain+spectra}a-b. 

The ROM harmonics can be considered as a perturbation in the reflected
signal as most of the pulse energy remains in the fundamental. On
the contrary, the CSE harmonics consume most of the laser pulse energy.
This is nicely seen in the spectral intensity of the reflected fundamental
for the both cases (compare Fig.~\ref{fig:time_domain+spectra}b
and d). As the absorption is negligible, the energy losses at the
fundamental frequency can be explained solely by the energy transfer
to high harmonics. We can roughly estimate this effect by $I_{0}^{BGP}/I_{0}^{CSE}\approx\int_{1}^{\infty}\omega^{-8/3}\, d\omega/\int_{1}^{\infty}\omega^{-4/3}\, d\omega=5$.
This value is quite close to the one from the PIC simulations: $I_{0}^{(\text{Fig. 1b})}/I_{0}^{(\text{Fig. 1d})}=3.7$. 

Further, we can estimate amplitude of the CSE attosecond pulse analytically
from the spectrum. Since the harmonic phases are locked, for an arbitrary
power law spectrum $I(\omega)\propto\omega^{-q}$ and a spectral filter
$\left(\omega_{\mathrm{low}},\omega_{{\scriptstyle \mathrm{high}}}\right)$
we integrate the amplitude spectrum and obtain: 

\begin{eqnarray}
E_{\text{atto}} & \approx & \frac{2\sqrt{\left.I\right|_{\omega=\omega_{1}}}}{q-2}\left[\left(\frac{\omega_{0}}{\omega_{\text{low}}}\right)^{\frac{q}{2}-1}-\left(\frac{\omega_{0}}{\omega_{\text{high}}}\right)^{\frac{q}{2}-1}\right]\label{eq:atto_amplitude}\end{eqnarray}
Apparently, when the harmonic spectrum is steep, i.e. \textit{$q>2$},
the radiation is dominated by the lower harmonics $\omega_{\mathrm{low}}$.
This is the case of the BGP spectrum \textit{$q=8/3$}. That is why
one needs a spectral filter to extract the attosecond pulses here.
The situation changes drastically for slowly decaying spectra with
\textit{$q<2$} like the CSE spectrum with $q=4/3$. In this case,
the radiation is dominated by the high harmonics $\omega_{{\scriptstyle \mathrm{high}}}$.
Even without any spectral filtering the radiation takes on the shape
of an attosecond pulse. As a rule of thumb formula for the attosecond
peak field of the unfiltered CSE radiation we can write: \begin{equation}
E_{\text{atto}}^{\text{CSE}}\approx\sqrt{3}\left(m_{c}^{1/3}-1\right)E_{0}\label{eq:atto_peak}\end{equation}
Using $m_{c}=\omega_{c}/\omega_{0}=225$, the lower of the two cut-off
harmonic numbers used for comparison with the PIC spectrum in Fig.~\ref{fig:time_domain+spectra}d,
we obtain $E_{\text{peak}}=8.8\, E_{0}$. This is in nice agreement
with Fig.~\ref{fig:time_domain+spectra}c.

\section{\label{sec:Parametrical-Dependence}Parametrical Dependence of Harmonics
Radiation}

Now, we have a look at the dependence of the harmonics radiation in
and close to the CSE regime on the laser and plasma parameters. Exemplary, 
the laser intensity and the preplasma scale length are varied here.
The pulse duration however will be left constantly short, so that we can simply 
focus our interest on the main optical cycle. For longer pulses, the extent of nanobunching may 
vary from one optical cycle to another, which makes a parametrical study more difficult.
 We are going to examine two dimensionless key quantities: 
the intensity boost $\eta\equiv\max(E_{r}^{2})/\max(E_{i}^{2})$ and the pulse compression
$\Gamma\equiv(\omega_{0}\tau)^{-1}$. It is straightforward to extract
both magnitudes from the PIC data, and both are quite telling. The
intensity boost $\eta$ is a sign of the mechanism of harmonics generation.
If the ARP boundary condition Eq.~\eqref{eq:ARP_boundary} is approximately
valid, we must of course have $\eta\approx1$. Then again, if the
radiation is generated by nanobunches, we expect to have strongly
pronounced attosecond peaks (compare Eq.~\eqref{eq:atto_peak}) in
the reflected radiation and therefore $\eta\gg1$. The pulse compression
$\Gamma$ is defined as the inverse of the attosecond pulse duration.
In the nanobunching regime, we expect it to be roughly proportional
to $\eta$, as the total efficiency of the attosecond pulse generation
remains $\eta_{\text{atto}}\lesssim1$, compare Eq.~\eqref{eq:cse_eff}.
In the BGP regime, there are no attosecond pulses observed without
spectral filtering. So the FWHM of the intensity peak is on the order
of a quarter laser period, and we expect $\Gamma\sim1$.

\begin{figure}
\includegraphics[width=1\columnwidth]{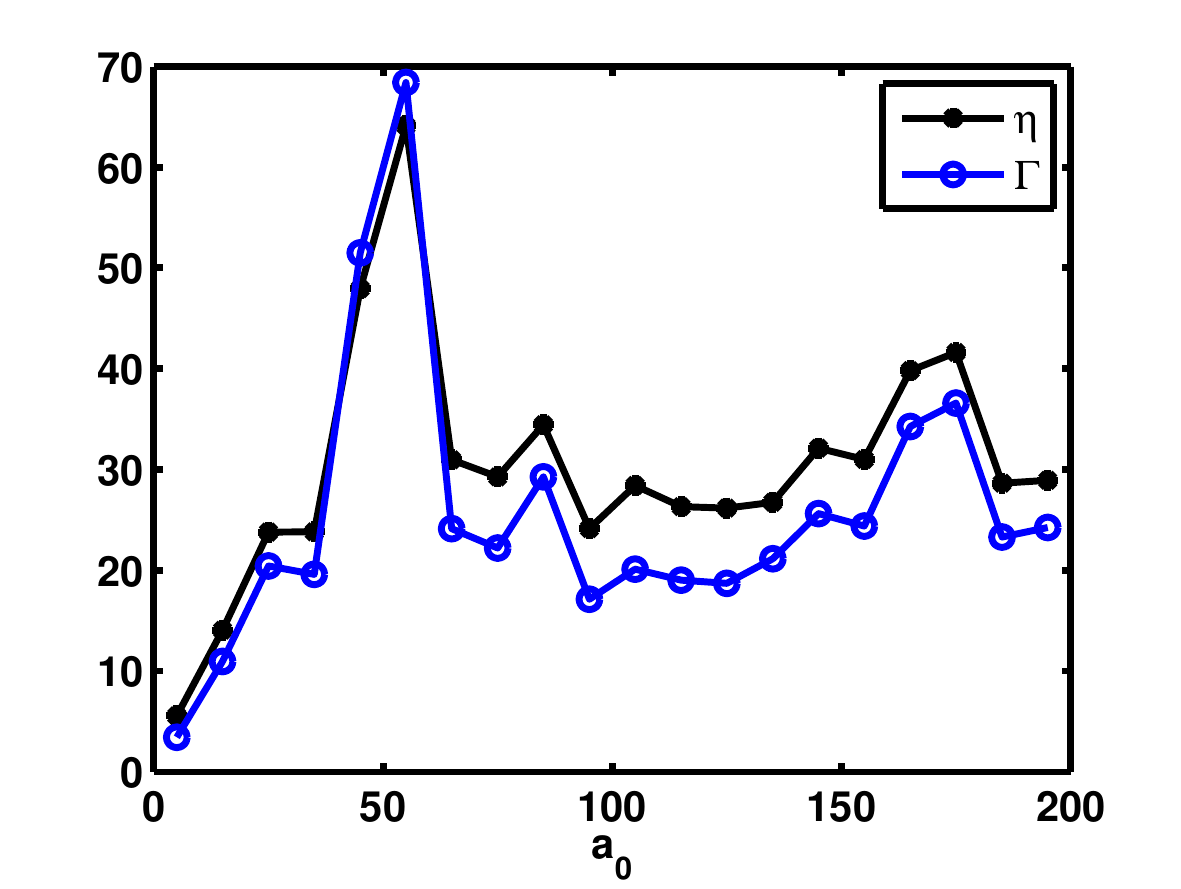}\caption{\label{fig:new_model_019_eta}Dependence of the intensity boost $\eta=\max(E_{r}^{2})/\max(E_{i}^{2})$
and the pulse compression $\Gamma=(\omega_{0}\tau)^{-1}$, where $\tau$
is the FWHM width of the attosecond intensity peak in the reflected
radiation, on $a_{0}$. The laser amplitude $a_{0}$ is varied between
5 and 195 in steps of 10. Other parameters are the same as in Fig.~\ref{fig:generation}b.}

\end{figure}

In figure~\ref{fig:new_model_019_eta} the two parameters $\eta$
and $\Gamma$ are shown in dependence of $a_{0}$. Except for the
variation of $a_{0}$, the parameters chosen are the same as in Figs.\ref{fig:time_domain+spectra}c-d,
\ref{fig:generation}b and \ref{fig:vy}.

First of all we notice, that for all simulations in this series with
$a_{0}\gg1$, we find $\eta\gg1$. Thus, Eq.~\eqref{eq:ARP_boundary}
is violated in all cases. Since we also notice $\Gamma\gg1$ and $\Gamma\sim\eta$,
we know, that the radiation is emitted in the shape of attosecond
peaks with an efficiency of the order 1. This indicates, that we can
describe the radiation as CSE. The perhaps most intriguing feature
of Fig.~\ref{fig:new_model_019_eta} is the strongly pronounced peak
of both curves around $a_{0}=55$. We think that because of some very
special phase matching between the turning point of the electron bunch
and of the electromagnetic wave, the electron bunch experiences an
unusually high compression at this parameter settings. This is the
case that was discussed in section~\ref{sec:New-Regime}.

\begin{figure}

\includegraphics[width=1\columnwidth]{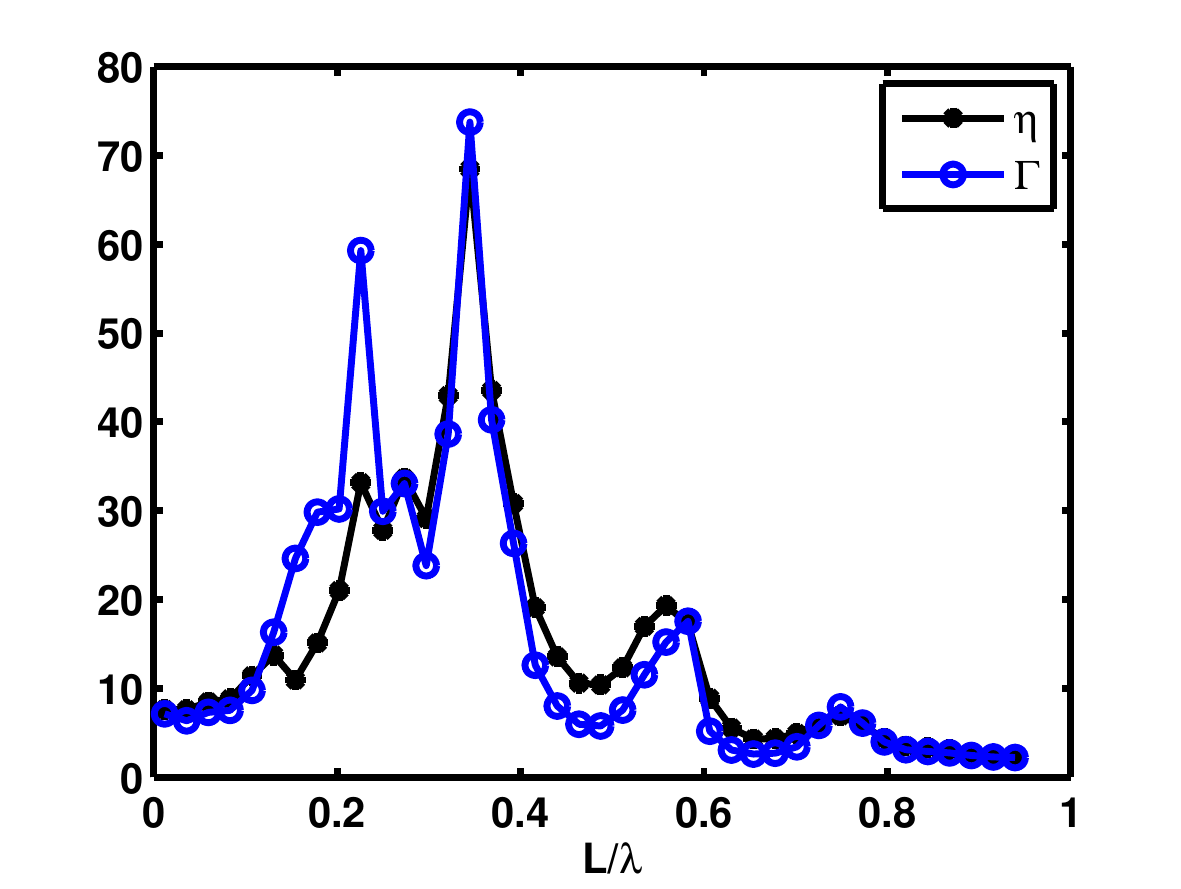}

\caption{\label{fig:new_model_024_eta}Dependence of the intensity boost $\eta=\max(E_{r}^{2})/\max(E_{i}^{2})$
and the pulse compression $\Gamma=(\omega_{0}\tau)^{-1}$, on the
plasma scale length in units of the laser wavelength $L/\lambda$
in the lab frame. Except for the plasma scale length, parameters are
the same as in Fig.~\ref{fig:generation}b. The plasma ramp is again
an exponential one $\propto\exp(x/L)$.}

\end{figure}

Figure~\ref{fig:new_model_024_eta} shows the two parameters $\eta$
and $\Gamma$ as functions of the plasma gradient scale length $L$.
It is seen that both functions possess several local maxima. Further,
$\eta$ and $\Gamma$ behave similar apart from one runaway value
at $L=0.225\lambda$, where the FWHM peak duration is extremely short,
but the intensity boost is not as high. A look at the actual field
data tells us that in this case the foot of the attosecond peak is
broader, consuming most of the energy. This deviation might e.g. be
caused by a different, non-gaussian shape of the electron nanobunch.

The maximum of both functions lies around $L=0.33\lambda$, the parameter
setting analyzed in detail before. In the limit of extremely small
scale lengths $L\lesssim0.1\lambda$, $\eta$ and $\Gamma$ become
smaller, but they remain clearly bigger than one. Thus the reflection
in this parameter range can still not very well be described by the
ARP boundary condition. For longer scale lengths $L>0.8\lambda$,
both key values approach 1, so the ARP boundary condition can be applied
here. This is a possible explanation, why the BGP spectrum \eqref{eq:ARP_spec}
could experimentally measured at oblique incidence \cite{Dromey2006}.

\section{Conclusions}

In this paper we have identified a novel mechanism of harmonics generation
at overdense plasma surfaces, leading to the flattest harmonics spectrum
known so far. Extremely dense and narrow peaks in the electron density,
we call them nanobunches, are responsible for the radiation. The spectrum
can be understood as a 1D synchrotron spectrum emitted by a relativistically
moving, extremely narrow and dense electron layer. Like the CWE and
ROM harmonics, the CSE harmonics are phase locked and appear in the
form of attosecond pulses. In contrast to CWE and ROM, the attosecond
pulses are visible immediately, therefore no energy needs to be wasted
in spectral filtering and the energy efficiency of attosecond pulse
generation is close to 100\%.

The work has been supported by DFG Transregio SFB TR18 and Graduiertenkolleg
GRK 2103.


\begin{thebibliography}{10}

\bibitem{PukhovNatPhys2006}
A.~Pukhov,
\newblock Nature Physics {\bf 2}, 439 (2006).

\bibitem{NomuraNatPhys2009}
Y.~Nomura, R.~Horlein, P.~Tzallas, B.~Dromey, S.~Rykovanov, Z.~Major,
  J.~Osterhoff, S.~Karsch, L.~Veisz, M.~Zepf, D.~Charalambidis, F.~Krausz, and
  G.~D. Tsakiris,
\newblock Nature Physics {\bf 5}, 124 (2009).

\bibitem{quere:125004}
F.~Quere, C.~Thaury, P.~Monot, S.~Dobosz, P.~Martin, J.-P. Geindre, and
  P.~Audebert,
\newblock Phys. Rev. Lett. {\bf 96}, 125004 (2006).

\bibitem{lichters:3425}
R.~Lichters, J.~M. ter Vehn, and A.~Pukhov,
\newblock Phys. Plasmas {\bf 3}, 3425 (1996).

\bibitem{Dromey2009}
B.~Dromey, D.~Adams, R.~Horlein, Y.~Nomura, S.~G. Rykovanov, D.~C. Carroll,
  P.~S. Foster, S.~Kar, K.~Markey, P.~McKenna, D.~Neely, M.~Geissler, G.~D.
  Tsakiris, and M.~Zepf,
\newblock Nature Physics {\bf 5}, 146 (2009).

\bibitem{tarasevitch:103902}
A.~Tarasevitch, K.~Lobov, C.~Wunsche, and D.~von~der Linde,
\newblock Phys. Rev. Lett. {\bf 98}, 103902 (2007).

\bibitem{bgp-theory2006}
T.~Baeva, S.~Gordienko, and A.~Pukhov,
\newblock Phys. Rev. E {\bf 74}, 046404 (2006).

\bibitem{Dromey2006}
B.~Dromey, M.~Zepf, A.~Gopal, K.~Lancaster, M.~S. Wei, K.~Krushelnick,
  M.~Tatarakis, N.~Vakakis, S.~Moustaizis, R.~Kodama, M.~Tampo, C.~Stoeckl,
  R.~Clarke, H.~Habara, D.~Neely, S.~Karsch, and P.~Norreys,
\newblock Nature Physics {\bf 2}, 456 (2006).

\bibitem{boyd:125004}
T.~J.~M. Boyd and R.~Ondarza-Rovira,
\newblock Phys. Rev. Lett. {\bf 101}, 125004 (2008).

\bibitem{Bourdier1983}
A.~{Bourdier},
\newblock Physics of Fluids {\bf 26}, 1804 (1983).

\bibitem{adBPuk2007}
D.~{An der Br{\"u}gge} and A.~{Pukhov},
\newblock Physics of Plasmas {\bf 14}, 093104 (2007).

\bibitem{PhysRevLett.93.115002}
S.~Gordienko, A.~Pukhov, O.~Shorokhov, and T.~Baeva,
\newblock Phys. Rev. Lett. {\bf 93}, 115002 (2004).

\bibitem{Jackson}
J.~D. Jackson,
\newblock {\em Classical {E}lectrodynamics},
\newblock J. Wiley \& Sons Inc., 3rd ed. edition, 1998.

\bibitem{1984Swallowtail}
J.~N.~L. {Connor}, P.~R. {Curtis}, and D.~{Farrelly},
\newblock Journal of Physics A Mathematical General {\bf 17}, 283 (1984).

\end{thebibliography}
\end{document}